%% file: 24509_ja.tex
\documentclass{aa}  

\usepackage{graphicx}
\usepackage{txfonts}
\bibpunct{(}{)}{;}{a}{}{,} 

\input{defs}  

\usepackage{marginnote}
\usepackage{color}
\begin{document}


\title{\kep\ detection of a new extreme planetary system \\
orbiting the subdwarf-B pulsator KIC~10001893
\subtitle{}
}

\author{R. Silvotti\inst{1}, S. Charpinet\inst{2,3}, E. Green\inst{4},
G. Fontaine\inst{5}, J.~H. Telting\inst{6}, R.~H.~\O stensen\inst{7},\\
V. Van Grootel\inst{8}, A.~S. Baran\inst{9}, S. Schuh\inst{10,11},
L. Fox Machado\inst{12}
%
%
%
}

\offprints{R. Silvotti}

\institute{INAF--Osservatorio Astrofisico di Torino, via Osservatorio
20, 10025 Pino Torinese, Italy\\
\email{silvotti@oato.inaf.it}
\and
Universit\'e de Toulouse, UPS-OMP, IRAP, 31400 Toulouse, France\\
\email{stephane.charpinet@irap.omp.eu}
\and
CNRS, IRAP, 14 Av. E. Belin, 31400 Toulouse, France
\and
Steward Observatory, University of Arizona, 933 North Cherry Avenue, Tucson, 
AZ, 85721, USA\\
\email{egreen@email.arizona.edu}
\and
D\'epartement de Physique, Universit\'e de Montr\'eal, C.P. 6128, 
Succ. Centre-Ville, Montr\'eal, Qu\'ebec H3C 3J7, Canada\\
\email{fontaine@astro.umontreal.ca}
\and
Nordic Optical Telescope, 
Rambla Jos\'e Ana Fern\'andez P\'erez 7, 38711 Bre\~na Baja, Spain\\
\email{jht@not.iac.es}
\and
Instituut voor Sterrenkunde, KU Leuven, Celestijnenlaan 200D, 3001 Leuven, 
Belgium\\
\email{roy@ster.kuleuven.be}
\and
Institut d'Astrophysique et de G\'eophysique, Universit\'e de Li\`ege, 
17 all\'ee du 6 Ao$\hat{\rm u}$t, 4000 Li\`ege, Belgium\\
\email{valerie.vangrootel@ulg.ac.be}
\and
Mt Suhora Observatory, Cracow Pedagogical University, ul. Podchorazych 2, 
30-084 Krakow, Poland\\
\email{sfbaran@cyf-kr.edu.pl}
\and
Institut f\"ur Astrophysik, Georg-August-Universit\"at G\"ottingen, 
Friedrich-Hund-Platz 1, 37077 G\"ottingen, Germany\\
\email{schuh@astro.physik.uni-goettingen.de}
\and
Max Planck Institute for Solar System Research, Max-Planck-Strasse 2,
37191 Katlenburg-Lindau, Germany
\and
Observatorio Astron\'omico Nacional, Universidad Nacional Aut\'onoma de 
M\'exico, BC, 22860, Ensenada, Mexico\\
\email{lfox@astrosen.unam.mx}
}

\date{Received .... 2014 / Accepted .....}

\abstract
{KIC~10001893 is one out of 19 subdwarf-B (sdB) pulsators observed by the 
\kep\ spacecraft in its primary mission.
In addition to tens of pulsation frequencies in the $g$-mode domain, its 
Fourier spectrum shows three weak peaks at very low frequencies, which is too 
low to be explained in terms of $g$ modes.
The most convincing explanation is that we are seeing the orbital modulation
of three Earth-size planets (or planetary remnants) in very tight orbits,
which are illuminated by the strong stellar radiation.
The orbital periods are P$_1$=5.273, P$_2$=7.807, and P$_3$=19.48 hours, and 
the period ratios P$_2$/P$_1$=1.481 and P$_3$/P$_2$=2.495 are very close to
the 3:2 and 5:2 resonances, respectively.
One of the main pulsation modes of the star at 210.68 \muHz\ corresponds to 
the third harmonic of the orbital frequency of the inner planet, suggesting 
that we see, for the first time in an sdB star, $g$-mode pulsations tidally 
excited by a planetary companion.
The extreme planetary system that emerges from the \kep\ data is very similar 
to the recent discovery of two Earth-size planets orbiting the sdB
pulsator KIC~05807616 \citep{charpinet11a}.}
%
%
%

\keywords{planetary systems -- stars: horizontal-branch -- stars: 
oscillations}

\authorrunning{R. Silvotti et al.}
\titlerunning{\kep\ detection of a new extreme planetary system orbiting
the subdwarf-B pulsator KIC~10001893}

\maketitle


\section{Introduction}

What happens to the planets, in particular to the inner planets, at the end of 
stellar evolution is largely unknown.
After the main sequence, dramatic changes in the planetary orbits and even the
complete evaporation of the inner planets can occur, following the 
red giant branch (RGB) and asymptotic giant branch (AGB) expansion 
of the host star
\citep{villaver07, villaver09}.
During the RGB or AGB expansion, the opposite effects of stellar mass loss and 
tidal interactions may determine a gap in the final distribution of orbital 
distances and periods \citep{nordhaus13}.
The first step in studying the effects of the RGB expansion is
to search for planetary systems on the horizontal branch (HB).
The star studied in this article is a subdwarf B (sdB), located on the 
extreme horizontal branch (EHB).
%

The sdB stars are a homogeneous class of very hot objects with a mass 
distribution peaked near 0.47 \msun\ \citep{fontaine12} and with a very 
thin hydrogen envelope (see the review by \citealt{heber09} for more details 
on hot subdwarfs).
Almost all the envelope was lost near the tip of the RGB.
Such strong envelope ejection is explained well in terms of close binary 
evolution for half of the sdBs that have a close stellar companion, generally 
an M dwarf or a white dwarf \citep{han02, han03}.
But it is more problematic for the other half of apparently single sdB stars.
The presence of close and massive planets or brown dwarfs (BDs) is a possible
explanation (\citealt{soker98}, \citealt{nelemans_tauris98}), that seems 
corroborated by recent calculations \citep{han12}.

The first detections of substellar companions to sdB stars started seven years 
ago, and now we know about thirteen sdB stars with 
planet/BD candidates belonging to three different groups that are quite 
distinct in terms of orbital distance and planetary mass. 
In order of decreasing orbital distance we find: 
i) eight planet/BD candidates in wide orbits (orbital periods between 3.2 and 
$\sim$16 yrs), with masses between $\sim$2 and $\sim$40 \mjup\
(\citealt{silvotti07}; \citealt{lee09} and \citealt{beuermann12b, qian09} and 
\citealt{beuermann12a, qian12}; \citealt{beuermann12a, lutz11} and 
\citealt{schuh14});
%
%
ii) two Earth-mass planet candidates around the sdB pulsator KIC~05807616, 
with orbital periods of 5.8 and 8.2 hours \citep{charpinet11a}; 
iii) at least two BD candidates with short orbital periods of a few hours 
(\citealt{geier11}, \citealt{schaffenroth14}, see also \citealt{geier12}). 
These three groups correspond to three different detection methods -- timing, 
illumination effects, and radial velocities (RVs), respectively --  that sample
three different regions of the $a-m_p$ (semi-major axis vs planetary mass) 
plane (see Fig.~1 of \citealt{silvotti14}).

The planet candidates presented in this paper belong to the second group and
closely resemble the planets of KIC~05807616.
These two planetary systems show very similar characteristics
(both with two or three Earth-size planets very close to the parent star
and very hot), suggesting that these extreme planetary systems around sdB 
stars might not be so rare and implying that planets could indeed play a role 
during the envelope ejection that is needed to form an sdB star.

%
%
%
%
%
%
%

The subject of this article has implications in different overlapping fields:
sdB evolution, common envelope (CE) ejection mechanisms, survival of a 
planet to a CE phase, and planetary system evolution in general.


\section{\kep\ data and ephemeris}

With a \kep\ magnitude of 15.85, KIC~10001893 is one out of 19 subdwarf B 
(sdB) pulsators observed by the \kep\ spacecraft in its primary mission
(\citealt{ostensen10,ostensen11,ostensen12}, \citealt{kawaler10a,kawaler10b}, 
\citealt{reed10,reed11,reed12}, \citealt{vangrootel10}, 
\citealt{baran11,baran12}, \citealt{charpinet11b}).
The spectroscopic atmospheric parameters of KIC~10001893, 
$\teff=26 \ 700\pm300$ K, $\logg=5.30\pm0.04$, and $\lheh=-2.09\pm0.1$ are 
very compatible with the $g$-mode instability strip \citep{ostensen11}.
The star was observed intensively with the \kep\ spacecraft \citep{borucki10}
in the framework of the \kep\ Asteroseismic Science Consortium (KASC, 
\citealt{gilliland10}), mostly in short cadence mode (SC, 58.85~s sampling), 
for 36 monthly runs (period Q3.2, where ``Q'' stands for quarter, and 
continuously from Q6 to Q17.2, when the spacecraft lost its second of four 
gyroscope-like reaction wheels).
Moreover, it was observed in long-cadence mode (LC, 29.4 min sampling)
for five more months (Q3.1, Q3.3, and Q5).
In our analysis we used all the available data, which means 993.8~d of SC data
and 147.9~d of LC data.
%
%
The data were downloaded from the \kep\ Asteroseismic Science Operations 
Center (KASOC) website~\footnote{http://kasoc.phys.au.dk/kasoc/}.

\begin{figure}[t]
\label{dft1}
%
%
\includegraphics[width=9.0cm]{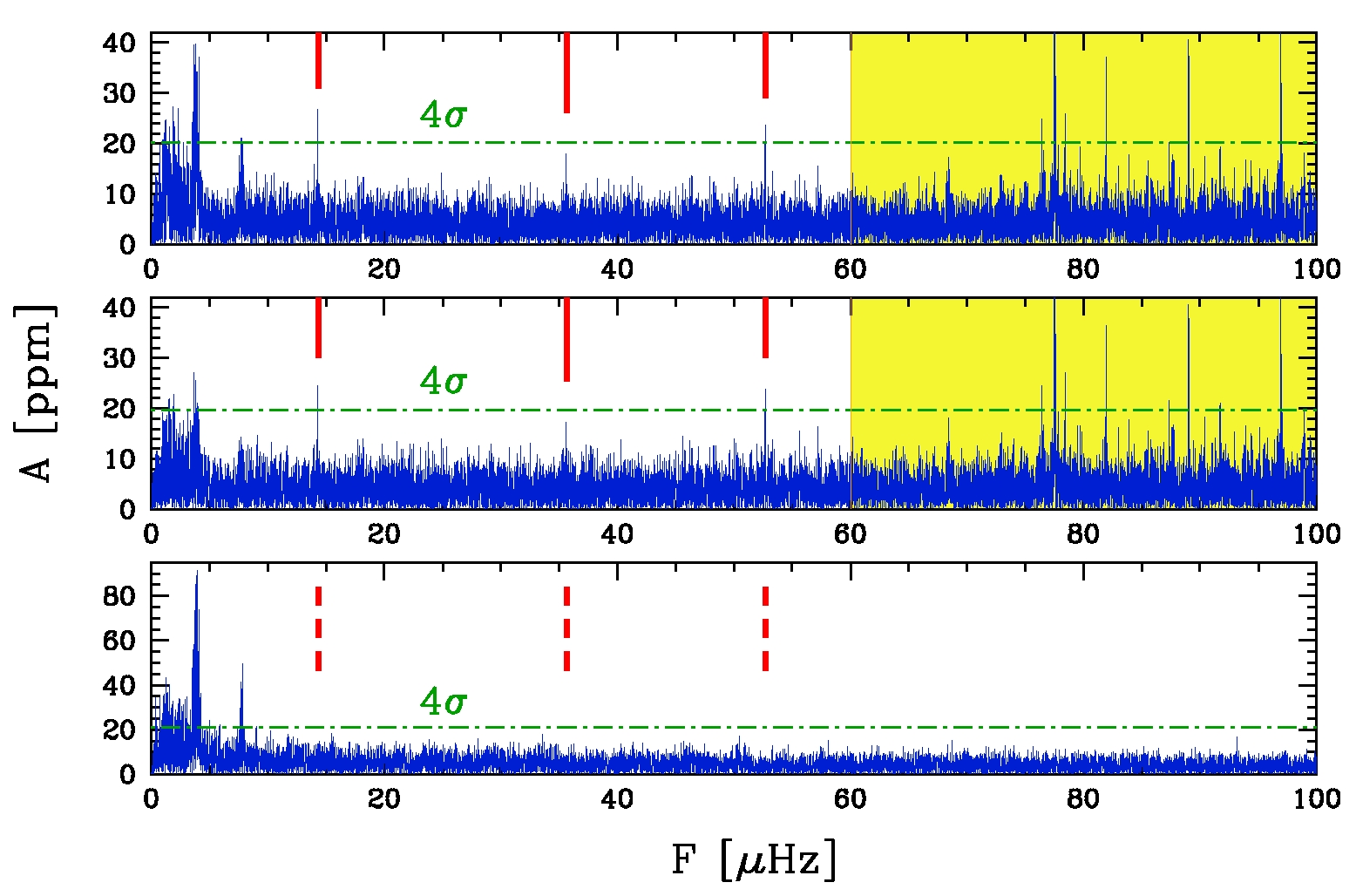}
\caption{Fourier transform of KIC~10001893 at low frequency.
The three panels show the amplitude spectrum using the standard
photometry produced by the \kep\ pipeline (top), the pixel photometry 
obtained optimizing the pixels on the target star (center), and the pixel 
photometry using only 1 pixel, centered on the nearby star (bottom).
The plot demonstrates that the frequency at about 4~\muHz\ and its first 
harmonic are caused by the nearby star, while the three other peaks 
(marked with a red vertical line) are intrinsic to KIC~10001893.
The shaded area at frequency higher than 60~\muHz\ (yellow in the electronic 
version) represents the region of the $g$-mode oscillations.}
\end{figure}

From a preliminary analysis using the photometric data produced by the \kep\ 
pipeline, the Fourier transform shows a rich spectrum with more than one 
hundred peaks, mostly concentrated between 76 and 441 \muHz.
At the low-frequency end of the amplitude spectrum, our attention was caught 
by a few peaks at very low frequency below 60 \muHz, that exceed the mean 
noise level by about four times (upper panel of Fig.~1).
These frequencies are too low to be produced by $g$-mode oscillations 
considering that the $\ell$=1 ($\ell$=2) atmospheric cut-off for the gravity 
modes is near 60 (105) \muHz\ for a star like KIC~10001893 
(\citealt{hansen85}, \citealt{charpinet11a}).
At lower frequencies, the reflective boundary condition at the surface is no 
longer valid.

To verify that these low-frequency peaks could be produced by a
nearby star located at approximately 8.5 arcsec northeast of KIC~10001893, 
we decided to use the pixel data that are available at the KASOC website and 
extract the photometry again for both the target and the close star, thereby 
optimizing the pixel combinations. 
The photometric contamination from nearby stars is a 
serious problem given the \kep\ pixel size of about 3.98 arcsec. 
For the close star we used only one pixel, the one corresponding to the PSF
maximum of this star, so that the contribution from KIC~10001893 was reduced 
to almost zero.
For the target we opted for a solution that could, at the same time, minimize 
the photons from the nearby star and also maximize the S/N: after 
different attempts, we decided to use, in each quarter, the pixel combination 
that was giving the maximum amplitude to the pulsation frequencies in the 
Fourier transform.
In Fig.~1 the amplitude spectrum of the two stars is shown.

We see very well that the peak at 4~\muHz\ and its harmonic near 8~\muHz\ are
produced by the nearby star (lower panel of Fig.~1).
Less clear is the origin of some power near 2~\muHz,\ but at $\sim$2~\muHz\ 
instrumental drifts may already have some importance.
On the other hand, it is clear from Fig.~1 that the three frequencies 
$f_1$=52.68~\muHz, $f_2$=35.58~\muHz,\ and $f_3$=14.26~\muHz, corresponding to 
the periods P$_1$=5.27~h, P$_2$=7.81~h, and P$_3$=19.48~h, are intrinsic to the
sdB star.
This conclusion is also confirmed from independent data processing using a 
model of the pixel response function determined from a combination of 
the \kep\ optical point-spread function and various systematics of the 
spacecraft, and this helps to avoid contamination from the neighboring stars.
When using all the available data, the amplitudes of the three orbital periods
have S/Ns of 4.8, 3.5, and 5.0 (Fig.~1).
At different amplitude levels, the three peaks are also visible when we 
consider shorter subsets containing only one third of the data (Fig.~2).
In the first subset, the peak at 35.58~\muHz, which was the faintest 
one in Fig.~1, reaches a S/N of 5.1 (upper panel of Fig.~2).
Most important, the phases of the three peaks are 
coherent in each independent subset.
The phase-folded light curves are shown in Fig.~3, and from the upper panel we 
note that P$_1$ is very close to four times one of the main pulsation periods 
of the star.

We have already seen that the low-frequency variations detected cannot be 
explained in terms of $g$ modes.
We add that they do not even correspond to linear combinations of the 
pulsation frequencies.\footnote{The maximum coincidence is for 
$f_1$=52.681~\muHz, so relatively close to the difference between two 
low-amplitude peaks: 204.670--152.018=52.652~\muHz\ (the formal resolution 
being 0.013~\muHz).
However, it would be very surprising that these faint peaks (37 and 67 ppm, 
respectively) give rise to linear combinations, while no combinations peaks 
are seen for much stronger pulsation frequencies.
On the other hand, $f_2$=35.582~\muHz\ is not very far from 
359.680--323.980=35.700~\muHz, two high-amplitude peaks (520 and 823 ppm, 
respectively), but here the difference in frequency is nine times the 
frequency resolution.}
%
%
Other explanations like surface spots are very unlikely.
Spots have never been observed in sdB stars, which have stable radiative 
envelopes (no convection) and are non-magnetic.
Indeed we do not see any signature of a magnetic field in our optical 
spectra or in the pulsation spectrum. 
Moreover, sdB stars have relatively long rotational periods, P$_{\rm ROT}> 1$ 
day from spectroscopy (\citealt{geier_heber12}), while values up to $\sim$40 
days are obtained from pulsation rotational splitting (e.g., 
\citealt{baran12}).
This rules out all (known) alternatives other than orbitally driven 
modulations.
In conclusion, the most convincing explanation for the three low-frequency
variations observed in the light curve of KIC~10001893 is the presence of
three low-mass bodies orbiting the star in tight orbits.

A strong argument in favor of this interpretation is given by the period 
ratios P$_2$/P$_1$=1.480 and P$_3$/P$_2$=2.495, which are very close to the 
3:2 and 5:2 resonances.
%
%
Another interesting aspect is that one of the main pulsation modes of the 
star, the one at 210.68 \muHz, corresponds to the third harmonic of the 
orbital frequency of the inner planet (52.68 $\times$ 4 = 210.72 \muHz), 
suggesting a tidal resonance (upper panel of Fig.~3).\footnote{
The period ratios close to resonances and the relation between P$_1$ and one 
of the main pulsation periods of the star allow to exclude also that the 
photometric variations that we see are produced by any instrumental effects.}
%
Although tidally excited pulsations have already been observed in other
stars (e.g., \citealt{hambleton13}), this is the first time that we have a
clear indication of this phenomenon in an sdB star.

Fitting the Kepler data with sinusoidal waves, it is possible to compute the 
ephemeris of the three planet candidates:

\vspace{-4mm}

\begin{equation}
BJD^{\rm TDB}_1 = 2455093.4293(53) + 0.2197(34) ~ E   
\end{equation}

\vspace{-7mm}

\begin{equation}
BJD^{\rm TDB}_2 = 2455093.325(44) + 0.3253(59) ~ E 
\end{equation}

\vspace{-7mm}

\begin{equation}
BJD^{\rm TDB}_3 = 2455093.506(41) + 0.8116(62) ~ E 
.\end{equation}

\noindent
These equations give the times of the maximum flux, when each planet is 
at the maximum distance from our solar system (phase 0 in Fig.~3).
The reference epochs correspond to the first time at phase 0 of the \kep\ 
data.
The relatively large errors in equations (1), (2), and (3), obtained from 
a Monte-Carlo simulations, are due to the extremely faint signals.
A word of caution is needed about the phases, which are not very 
reliable.
From various tests, we have seen that the phases can change significantly
when we change the number of pulsation frequencies included in the fit. 
The phases reported here were computed using 60 pulsation frequencies with 
amplitudes greater than 20 ppm.

Unfortunately, despite the very high quality of the \kep\ data, it is not 
possible to confirm the presence of the three planets by an independent method
using the same data.
All the methods that make use of the R\"omer delay on pulsation timing --
O-C method \citep{silvotti07}, light curve fitting \citep{telting12}, and
orbital aliases in the Fourier domain (\citealt{telting12}, 
\citealt{shibahashi12}) -- require larger orbits.
%
%
%
%
With these short orbital periods,
%
the maximum light-travel time shift is only
($1.8 \times 10^{-5}\ M_{\mathrm {PE}}\ {\rm sin}~i$),
($2.3 \times 10^{-5}\ M_{\mathrm {PE}}\ {\rm sin}~i$), and 
($4.2 \times 10^{-5}\ M_{\mathrm {PE}}\ {\rm sin}~i$) seconds (with the planet 
mass $M_{\mathrm {PE}}$ in Earth masses, where
$i$ is the inclination of the system with respect to the line of sight).
And the amplitude of the first orbital sidelobes at $\pm \nu_{\mathrm {orb}}$ 
from the main pulsation frequency at 274.302~\muHz\ (which has an amplitude of 
0.1\%) would be
($3.0 \times 10^{-5}\ M_{\mathrm {PE}}\ {\rm sin}~i$),
($4.0 \times 10^{-5}\ M_{\mathrm {PE}}\ {\rm sin}~i$), and
($7.3 \times 10^{-5}\ M_{\mathrm {PE}}\ {\rm sin}~i$) ppm.
These numbers are well below the detection limit of \kep\ for any realistic 
planetary mass.


\begin{figure}[ht]
\label{dft2}
\includegraphics[width=9.0cm]{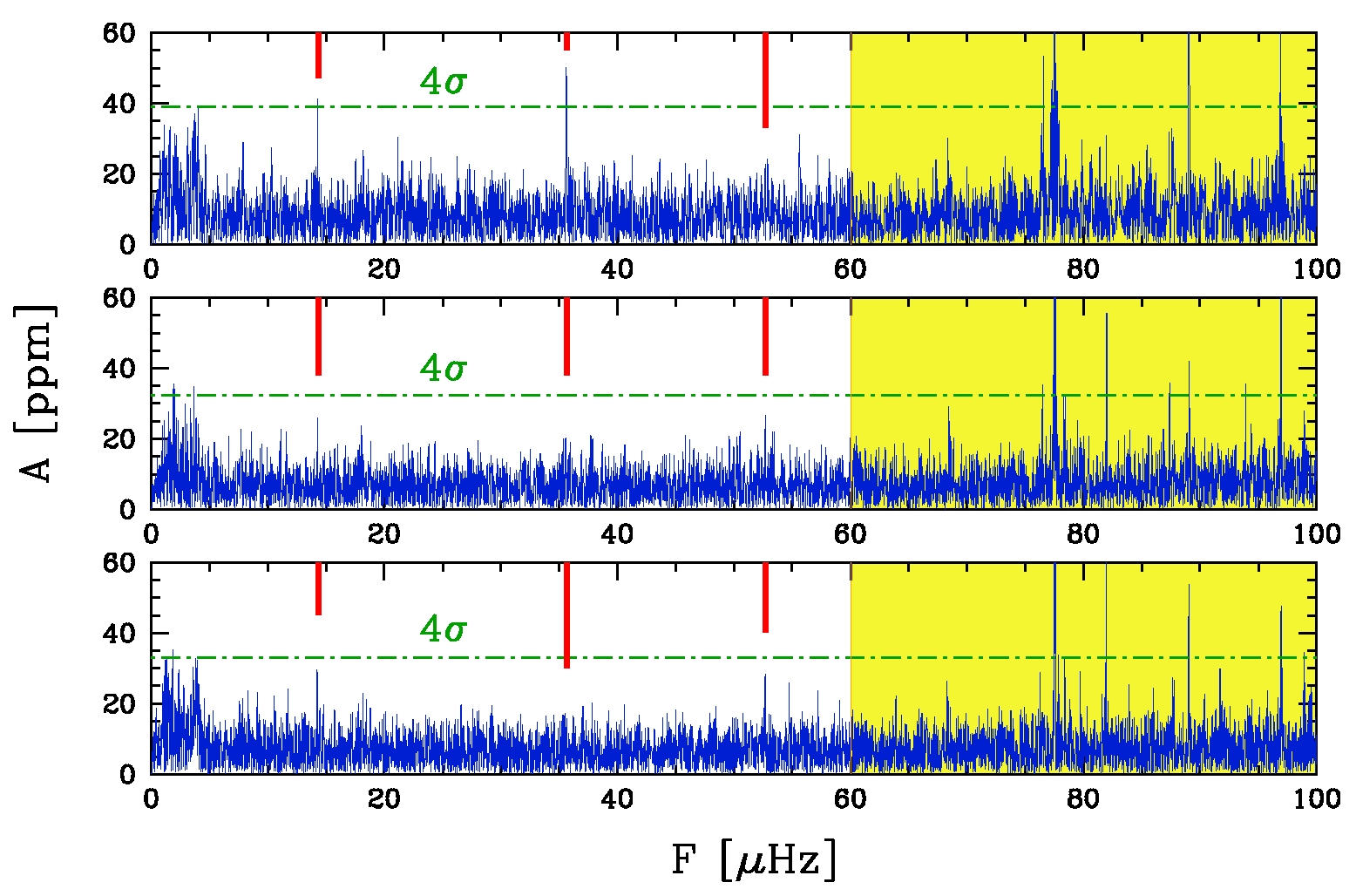}
\caption{Fourier transform of KIC~10001893 using three sets of independent
pixel photometry data with a duration of about 13-14 months each. From top to 
bottom: Q3+Q5+Q6+Q7+Q8.1, Q8.2+Q8.3+Q9+Q10+Q11+Q12.1+Q12.2, 
Q12.3+Q13+Q14+Q15+Q16+Q17.1+Q17.2.
The apparent changes in amplitude of the low-frequency peaks are mostly
consistent with statistical noise.}
%
%
\end{figure}

\begin{figure}[h]
\label{phase}
%
%
\includegraphics[width=8.8cm]{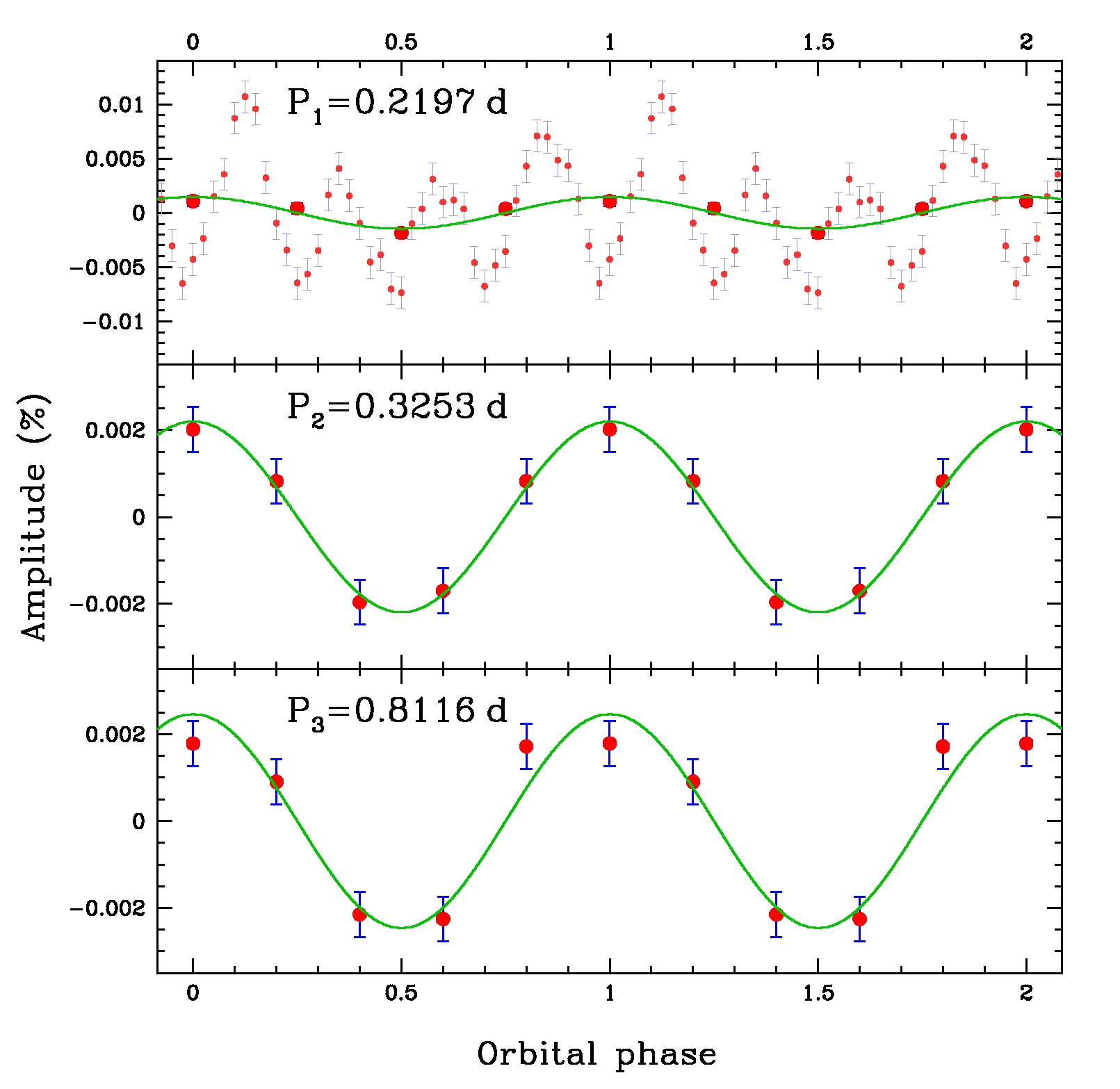}
\caption{Phase-folded light curves for the three planet candidates
and our best fits with a sine wave.
Each point is the mean flux in that phase bin.
The upper panel shows that the inner planet has a period very close to 4 
times one of the main pulsation periods of the star.
%
%
The error bars are obtained from the original flux uncertainties given by the 
\kep\ pipeline: $\sigma = \frac{1}{n} ~ (\sum_{i=1}^n \sigma_i^2)^{1/2}$.
%
%
}
\end{figure}


\section{Spectroscopy and radial velocities}

KIC~10001893 was observed in 2010, 2011, 2012, and 2013 using the 
William Herschel, the Nordic Optical, the Multi Mirror, and the 2.1~m
San Pedro Martir Telescopes (WHT, NOT, MMT, and UNAM) to measure 
possible radial velocity (RV) variations (Fig.~4).
\begin{figure}[ht]
\label{RVs}
\includegraphics[width=9.0cm]{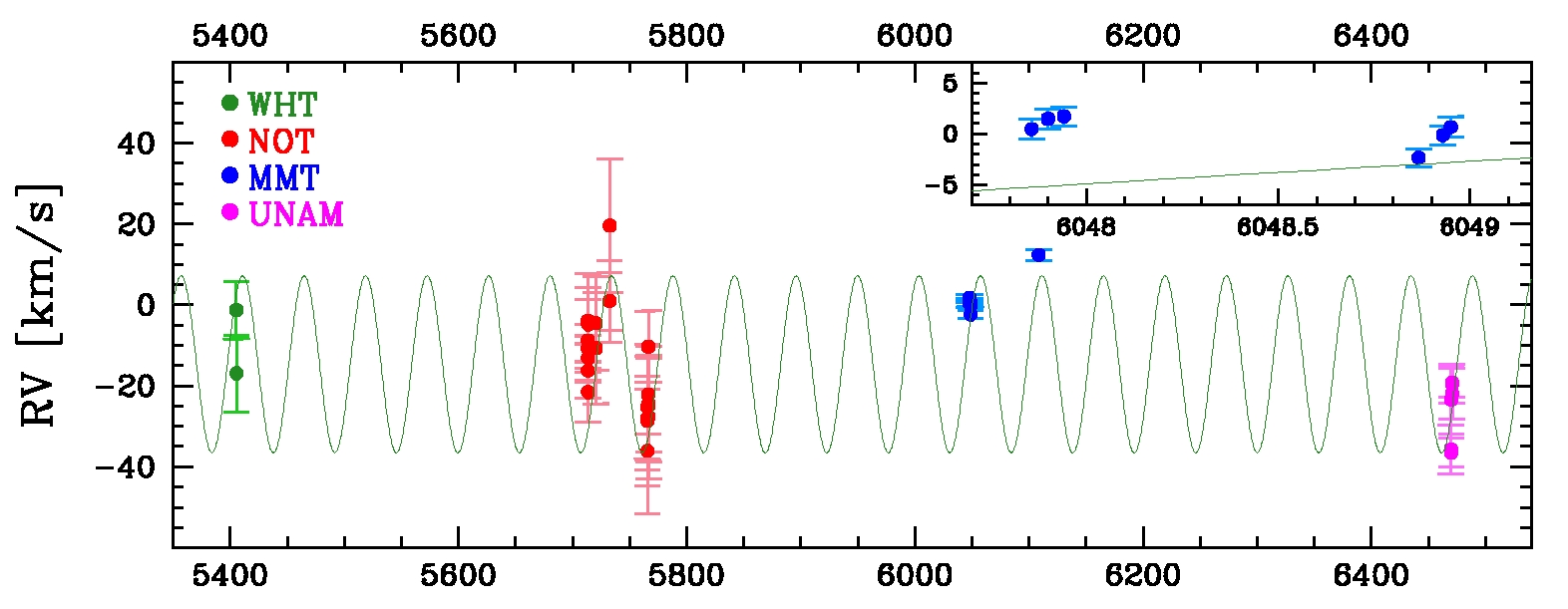}
\caption{adial velocities of KIC~10001893. The inset shows in detail the
MMT measurements of April 30 and May 1, 2012.
A tentative fit with a period of about 54 days is shown.}
\end{figure}

The six best RV measurements from the MMT on April 30 and May 1, 2012
are consistent with a constant radial velocity of $0.3 \pm 1.6$ km/s.
However, a single MMT spectrum taken two months later on June 30 indicates a
RV of $12.3 \pm 1.3$ km/s.
We do not have any reason to think that this last measurement has any problem, 
because spectra taken on the same night on another target, Feige~48, agree to 
better than 1 km/s with the binary orbit of Feige~48.
Unfortunately, we did not have the possibility to observe KIC~10001893 again
with MMT in 2012, 2013, or 2014.
Our conclusion is that with the RV data that we have we can exclude any RV 
amplitude larger than about 25 km/s.
Considering only the MMT data (last spectrum included), the amplitude must be 
at least 6 km/s.
If we associate an RV variation of this amplitude with the longer orbital 
period that we found in the \kep\ data (19.5 hours), the minimum mass of the 
companion (at $i=90^{\circ}$) would be about 17 Jupiter masses, which is 
enough to make the system dynamically unstable.
Thus, if the RV variation is real, we should instead think of a more massive 
faint (stellar) companion in a wide orbit. \footnote{The absence of Doppler 
beaming and ellipsoidal deformation of the sdB star also allow excluding that 
the photometric variations that we see are caused by a white dwarf companion 
in a close orbit.}
A tentative fit of the RV data with a period of about 54 days is shown in
Fig.~4.

With the many spectra that we got on KIC~10001893 (including some spectra
from the Bok telescope), it was possible to improve the atmospheric parameters 
of the star.
The WHT and NOT spectra were fit using LTE models with solar abundances
\citep{heber00}, while NLTE models 
were used to fit the Bok and MMT spectra.
Our best estimate is
$\teff=27 \ 500\pm500$, $\logg=5.35\pm0.05$ and $\lheh=-2.95\pm0.04$




%


\section{An extreme planetary system}

In this section we explore the properties of the three low-mass bodies 
according with the variations observed in the light curve.
As discussed by \citet{charpinet11a}, the light modulation that we see can be 
caused by the reflected light from the illuminated side of each low-mass 
companion.
The intensity of this effect depends on the Bond albedo $\rm{\alpha_B}$.
Another effect, which can be even more important for very hot stars and very 
close planets, is the thermal emission from each planet, which is modulated 
by the temperature difference between the heated day side and the cooler 
night-side hemisphere, assuming that the rotation of each planet is tidally 
synchronized to its orbital motion (a likely situation given the short 
orbital periods).
Under the assumption of radiative equilibrium and black-body re-emission, 
we introduce a second parameter $\rm{\beta}$, defined as the ratio between 
the two temperatures.
The equilibrium temperatures on the two hemispheres are given by

\vspace{-4mm}

\begin{equation}
T_{\rm {day}} = \left[\frac{1 - \alpha_{\rm B}}{2 ~ 
(1 + \beta^4)}\right]^{1/4} (R_{\star} / a)^{1/2} ~ \teff
\end{equation}

\vspace{-4mm}

\begin{equation}
T_{\rm night} = \beta ~ T_{\rm {day}}
\end{equation}

\noindent
where $a = [G M_{\star} / (4 \pi^2)]^{1/3} ~ P_{\rm orb}^{2/3}$ is the 
orbital separation (assuming $M_{\rm P} << M_{\star}$), $M_P$ is the planet 
mass, $R_{\star} = (G M_{\star} / g)^{1/2}$ and \teff\ 
are the radius and the spectroscopic effective temperature of the parent star.
The surface gravity $g$ is obtained from spectroscopy, and we assumed 
a canonical stellar mass $M_{\star}$=~0.47 \msun.

Then we computed the planetary radii and masses as a function of the 
parameters mentioned above and of the inclination, using
the following equations (see \citealt{charpinet11a}, Supplementary 
Information, for more details):

\vspace{-4mm}

\begin{equation}
R_{\rm P} = \left(\frac{A}{sin~{\it i}}\right)^{1/2} 
\left[\frac{\alpha_{\rm B}}{8 a^2} + \frac{1}{2 R_{\star}^2} ~ 
\frac{F(T_{\mathrm {day}}) - 
F(T_{\mathrm {night}})}{F(T_{\mathrm {eff}})}\right]^{-1/2}
\end{equation}

\vspace{-4mm}

\begin{equation}
F(T) = \int B_{\lambda}(T) ~ \epsilon_{\lambda}^{K} ~ d\lambda
\end{equation}

\vspace{-4mm}

\begin{equation}
M_{\rm P} = \frac{4}{3} ~ \pi ~ R_{\rm P}^3 ~ \bar{\rho}
\end{equation}

\noindent
in which $A$ is the measured (semi-)amplitude of the photometric 
modulation, $F(T)$ is the black body radiative flux within the \kep\
response function $\epsilon_{\lambda}^{K}$, $B_{\lambda}$ is the 
Planck distribution, and $\bar{\rho}$ is the mean planet density.

Finally we computed the RV amplitude $K$ of the parent star:

\vspace{-4mm}

\begin{equation}
K = \left( \frac{2\pi~G}{P} \right)^{1/3} \frac{M_{\rm P} ~ {\rm sin}~{\it i}}
{(M_{\star} + M_{\rm P})^{2/3}}
.\end{equation}

\noindent
The results of our analysis are summarized in Fig.~5.
We see that for any reasonable value of $\alpha_{\rm B}$ and 
$\beta$ (and for any inclination $\ga$ 1 degree), radii and masses are in 
the planetary range. 
Only with a very high albedo and a very small temperature difference between
the two hemispheres (isothermal surface) can the radii be larger than the 
Jupiter radius if the inclination is very low.
But these conditions, similar to Venus, would require a dense atmosphere, 
which is very difficult to imagine in an extremely hot planet in a close orbit 
that has presumably entered the star envelope in the past.
Even in the case of second-generation planets, formed after the RGB phase, it
is unlikely that a dense atmosphere can be kept in such extreme conditions.
Assuming $\alpha_{\rm B}$=0.1 and $\beta$=0.2, the radii of the planets are
smaller than Neptune's radius for almost any inclination while, for 
$i>14^{\circ} ~ (16^{\circ})$, 
planet 1 (2) has a radius smaller than the Earth.

\begin{figure*}
\centering
\includegraphics[width=17cm]{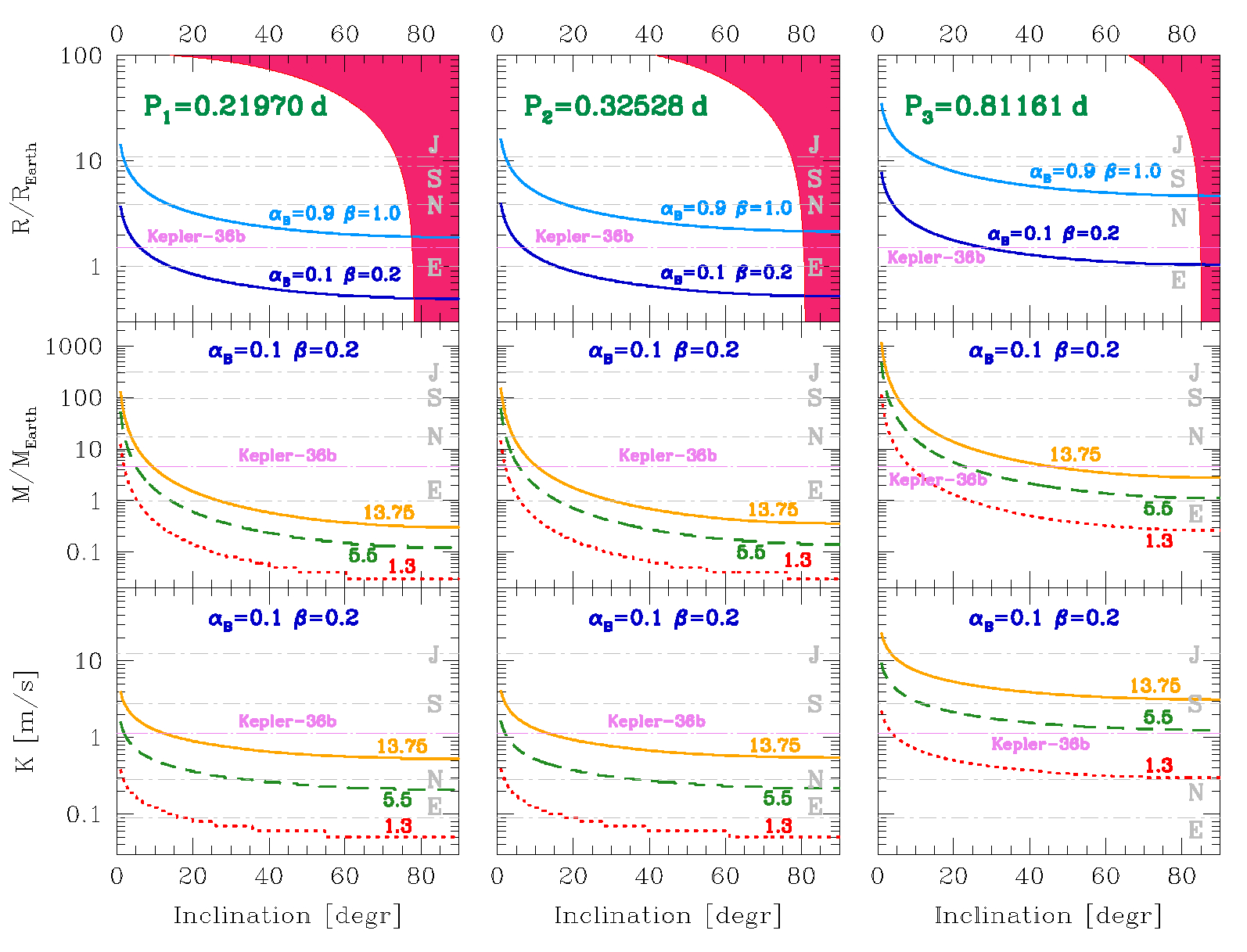}
\caption{Radii and masses of the three planets and host-star-projected RV 
amplitudes as functions of the inclination. We assume a Bond albedo 
$\alpha_{\rm B}$=0.10 and an average temperature contrast between night and 
day sides $\beta$=0.2; that is, approximately the values observed on Mercury. 
Only in the upper panels do we also show the opposite (less likely) extreme 
case with $\alpha_{\rm B}$=0.90 and $\beta$=1.0 (isothermal surface).
%
%
The red regions in the upper panels represent excluded domains due to the 
absence of eclipses.
Radii, masses, and stellar RV amplitudes for Jupiter, Saturn, Neptune, the 
Earth, and Kepler-36b, which has an estimated mean density of 7.5 g/cm$^3$
\citep{carter12}, are indicated for comparison.
The central and lower panels show three different curves with 
different mean densities: 1.3 (Jupiter), 5.5 (Earth), and 13.75 g/cm$^3$ 
(2.5 $\times$ Earth, corresponding to the mean density of a 100\% iron planet 
with Earth radius, \citealt{valencial10}, Fig.~4).
%
%
If KIC~10001893~b,c,d are the remnants of one or more massive planets,
%
%
they could have a high density corresponding to the iron-rich cores of their
Jovian progenitors.
%
}
\label{planets}
\end{figure*}


\section{Stability of the orbits vs inclination of the system}

From preliminary stability computations, assuming circular coplanar orbits,
and assuming that the three planets have same mean density, we see that the
stability of the orbits becomes critical in relatively short times
($<$10$^6$ yrs) when $\bar{\rho} ~ (sin~{\it i})^{-3/2}~\gsim 1000$ g/cm$^3$.
%
With mean densities of 1.3, 5.5, and 13.75 g/cm$^3$, the system is dynamically
stable only if the inclination is greater than about 0.7$^{\circ}$, 
1.8$^{\circ}$, and 3.3$^{\circ}$, respectively.
This constraint may help to understand a peculiar characteristic of the 
amplitude spectrum of this star, in which we do not see any rotational 
splitting of the pulsation frequencies (unlike all the other sdB pulsators 
observed by \kep).
The absence of multiplets may be due to a very low inclination 
(but higher than $0.7^{\circ}$ to $3.3^{\circ}$)
and/or to an extremely long rotation period.



\section{Discussion}

The planetary system described in this paper is a new candidate in the 
short list of systems with substellar companions in close orbits around 
evolved stars.
Such systems are crucial for studying the survivability of planets during the 
post-MS evolution of the host star, a question that has attracted some 
interest over the past few years
(\citealt{villaver07,villaver09}; \citealt{nordhaus10,passy12,spiegel12}; 
\citealt{mustill12,nordhaus13}).
As in the case of KIC~05807616, it is likely that the three planets of 
KIC~10001893 have lost a significant fraction of their mass during the CE 
phase.
Following \citet{passy12}, their initial mass, before the plunge-in, was 
most likely a few Jupiter masses ($M_{\rm J}$).
Much lower values should be excluded since they would cause the drag 
energy to be higher than the binding energy and the planet would be 
destroyed.
On the other hand, much higher values should also be excluded since, 
already at 10 $M_{\rm J}$, a companion would not be significantly affected 
by the CE phase.

Whether a few Jupiter masses are sufficient to remove
the entire envelope of the host star is not clear.
Following \citet[Fig.~2]{nordhaus13}, with orbital periods of 5, 8, and 19
hours, the minimum mass to eject the CE is about 13, 20, and 35 $M_{\rm J}$, 
respectively, assuming that all the orbital energy is used to unbind the CE 
($\alpha_{\rm CE}$=1).
Following \citet{han12}, when we approach the peak distribution of sdB masses 
at $\sim$0.47 \msun\ \citep{fontaine12}, a low-mass substellar companion may 
be sufficient to eject the common envelope, even when only the orbital energy 
released during the spiral-in is considered. 
Adding an extra term due to thermal energy, the minimum companion mass to 
eject the CE is reduced further.
An alternative scenario has been proposed by \citet{bear12} for KIC~05807616: 
a single massive planet ($M \ga 5 M_{\rm J}$), after having removed most of 
the stellar envelope, reaches the tidal-destruction radius at about 1 \rsun\ 
from the stellar core. Here the planet's gaseous envelope is removed, and the 
metallic core is disrupted in two or more Earth-size fragments that survive 
within the gaseous disk and migrate to resonant orbits (3:2 resonance in the 
case of KIC~05807616; 5:2 and 3:2 for KIC~10001893).
According to the same authors, the survival of the planets to the high 
evaporation rate due to the stellar UV radiation could be guaranteed by a 
planetary magnetic field ten times as strong as that of the Earth, which would 
substantially reduce the evaporation rate by holding the ionized gas.

Finally we want to add a few words of caution.
Although the most convincing interpretation of the low frequencies detected by
\kep\ in KIC~10001893 (and KIC~05807616) is the presence of Earth-size 
close-in planets, we do not consider these data and their analysis adequate 
to definitively prove that these extreme planetary systems around sdB 
stars do exist.
Rather, we think that these data are enough to seriously pose the question.
A final answer to this question will be given when it will be possible to 
confirm these detections with independent methods (e.g., RVs with PEPSI@LBT)
and/or find further planetary systems similar to them.
%
%
A program to search for planets/planetary remnants close to bright sdB stars
has recently been started using Harps-N@TNG \citep{silvotti14}.


\begin{acknowledgements}

The authors gratefully acknowledge the \kep\ team and everybody who
has contributed to making this mission possible.
Funding for the \kepmi\ is provided by NASA's Science Mission Directorate.
RS was supported by the PRIN-INAF on ``Asteroseismology: looking inside
the stars with space- and ground-based observations'' (PI Ennio Poretti).
LFM acknowledges the financial support from the UNAM under grant PAPIIT 
104612.
ASB gratefully acknowledges a financial support from the Polish National 
Science Center under project UMO-2011/03/D/ST9/01914.
VVG is an FNRS Research Associate.

%

\end{acknowledgements}


\bibliographystyle{aa} 


\end{document}

%% file: defs.tex
\newcommand{\kep}{{\em Kepler}}

\newcommand{\kepmi}{{\em Kepler Mission}}

\newcommand{\teff}{\ensuremath{T_{\mathrm{eff}}}}
\newcommand{\logg}{\ensuremath{\log g}}

\newcommand{\lheh}{\ensuremath{\log \left(\rm N_{\mathrm{He}}/\rm N_{\mathrm{H}}\right)}}

\newcommand{\gsim}{\raisebox{-1ex}{$\stackrel{{\displaystyle>}}{\sim}$}}

\newcommand{\msun}{${\mathrm{M}}_{\odot}$} 
\newcommand{\rsun}{${\mathrm{R}}_{\odot}$} 
\newcommand{\muHz}{$\mu$Hz}

\newcommand{\mjup}{$\rm M_{\rm Jup}$}

\def\mnras{MNRAS}
\def\aj{AJ}%
\def\apj{ApJ}%
\def\apjl{ApJ}%
\def\aap{A\&A}%
\def\pasp{PASP}%

%% file: 24509_ja.bbl
\begin{thebibliography}{}

\bibitem[\protect\citeauthoryear
{Baran et al.}{2011}]{baran11} 
Baran, A.~S., Kawaler, S.~D., Reed, M.~D., et al., 2011, \mnras, 414, 2871

\bibitem[\protect\citeauthoryear
{Baran et al.}{2012}]{baran12} 
Baran, A.~S., Reed, M.~D., Stello, D., et al., 2012, \mnras, 424, 2686

\bibitem[\protect\citeauthoryear
{Bear \& Soker}{2012}]{bear12} 
Bear, E., Soker, N., 2012, \apj, 749, L14

\bibitem[\protect\citeauthoryear
{Beuermann et al.}{2012a}]{beuermann12a} 
Beuermann, K., Breitenstein, P., Bski, B.~D., et al., 2012a, \aap, 540, A8

\bibitem[\protect\citeauthoryear
{Beuermann et al.}{2012b}]{beuermann12b} 
Beuermann, K., Dreizler, S., Hessman, F.~V., Deller, J., 2012b, \aap, 543, A138

\bibitem[\protect\citeauthoryear
{{Borucki} et al.}{2010}]{borucki10}
Borucki, W.~J., Koch, D., Basri, G., et al., 2010, Sci, 327, 977


\bibitem[\protect\citeauthoryear
{Carter et al.}{2012}]{carter12} 
Carter, J.~A., Agol, E., Chaplin, W.~J., et al., 2012, Science, 337, 556

\bibitem[\protect\citeauthoryear
{Charpinet et al.}{2011a}]{charpinet11a}
Charpinet, S., Fontaine, G., Brassard, P., et al., 2011a, Nature, 480, 496

\bibitem[\protect\citeauthoryear
{Charpinet et al.}{2011b}]{charpinet11b}
Charpinet, S., Van Grootel, V., Fontaine, G., et al., 2011b, \aap, 530, A3


\bibitem[\protect\citeauthoryear
{Fontaine et al.}{2012}]{fontaine12}
Fontaine, G., Brassard, P., Charpinet, S., et al., 2012, \aap, 539, A12

\bibitem[\protect\citeauthoryear
{Geier et al.}{2011}]{geier11} 
Geier, S., Schaffenroth, V., Drechsel, H., et al., 2011, \apj, 731, L22 

\bibitem[\protect\citeauthoryear
{Geier et al.}{2012}]{geier12} 
Geier, S., Classen, L., Br\"unner, P., et al., 2012, ASP Conf. Series, 452, 153


\bibitem[\protect\citeauthoryear
{Geier \& Heber}{2012}]{geier_heber12} 
Geier, S., Heber, U., 2012, \aap, 543, A149

\bibitem[\protect\citeauthoryear
{{Gilliland} et al.}{2010}]{gilliland10}
Gilliland, R.~L., Brown, T.~M., Christensen-Dalsgaard, J., et al. 2010, \pasp, 
122, 131



\bibitem[\protect\citeauthoryear
{Hambleton et al.}{2013}]{hambleton13} 
Hambleton K.~M., Kurtz D.~W., Prs\v{a} A., et al., 2013, \mnras, 434, 925

\bibitem[\protect\citeauthoryear
{Han et al.}{2002}]{han02} 
Han, Z., Podsiadlowski, Ph., Maxted, P.~F.~L., Marsh, T.~R., Ivanova, N., 
2002, \mnras, 336, 449

\bibitem[\protect\citeauthoryear
{Han et al.}{2003}]{han03} 
Han, Z., Podsiadlowski, Ph., Maxted, P.~F.~L., Marsh, T.~R., 2003, \mnras, 341,
669

\bibitem[\protect\citeauthoryear
{Han et al.}{2012}]{han12} 
Han, Z., Chen, X., Lei, Z., Podsiadlowski, P., 2012, PASP Conf. Series, 452, 3

\bibitem[\protect\citeauthoryear
{Hansen et al.}{1985}]{hansen85}
Hansen, C.~J., Winget, D.~E., Kawaler, S.~D., 1985, \apj, 297, 544

\bibitem[\protect\citeauthoryear
{Heber et al.}{2000}]{heber00} 
Heber, U.,  Reid, I.~N., Werner, K., 2000, \aap, 363, 198

\bibitem[\protect\citeauthoryear
{Heber}{2009}]{heber09} 
Heber, U., 2009, ARA\&A, 47, 211


\bibitem[\protect\citeauthoryear
{Kawaler et al.}{2010a}]{kawaler10a} 
Kawaler, S.~D., Reed, M.~D., Quint, A., et al., 2010a, \mnras, 409, 1487

\bibitem[\protect\citeauthoryear
{Kawaler et al.}{2010b}]{kawaler10b} 
Kawaler, S.~D., Reed, M.~D., \O stensen R.~H., et al., 2010b, \mnras, 409, 1509

\bibitem[\protect\citeauthoryear
{Lee et al.}{2009}]{lee09} 
Lee, J.~W., Kim, S.-L., Kim, C.-H., et al., 2009, \aj, 137, 3181

\bibitem[\protect\citeauthoryear
{Lutz}{2011}]{lutz11} 
Lutz, R., 2011, PhD thesis, Georg-August-Universit\"at G\"ottingen,
({\rm http://webdoc.sub.gwdg.de/diss/2011/lutz/lutz.pdf})



\bibitem[\protect\citeauthoryear
{Mustill \& Villaver}{2012}]{mustill12}
Mustill, A.~J., Villaver E., 2012, \apj, 761, 121

\bibitem[\protect\citeauthoryear
{Nelemans \& Tauris}{1998}]{nelemans_tauris98}
Nelemans, G., Tauris, T.~M., 1998, \aap, 335, L85

\bibitem[\protect\citeauthoryear
{Nordhaus et al.}{2010}]{nordhaus10} 
Nordhaus, J., Spiegel, D.~S., Ibgui, L., Goodman, J., Burrows, A., 2010,
\mnras, 408, 631

\bibitem[\protect\citeauthoryear
{Nordhaus \& Spiegel}{2013}]{nordhaus13} 
Nordhaus, J., Spiegel, D.~S., 2013, \mnras, 432, 500


\bibitem[\protect\citeauthoryear
{{\O}stensen et al.}{2010}]{ostensen10} 
{\O}stensen R.~H., Silvotti, R., Charpinet, S., et al., 2010, \mnras, 409, 1470

\bibitem[\protect\citeauthoryear
{{\O}stensen et al.}{2011}]{ostensen11} 
{\O}stensen, R.~H., Silvotti, R., Charpinet S., et al., 2011, \mnras, 414, 2860

\bibitem[\protect\citeauthoryear
{{\O}stensen et al.}{2012}]{ostensen12} 
{\O}stensen, R.~H., Degroote, P., Telting, J.~H., et al., 2012, \apj, 753, L17

\bibitem[\protect\citeauthoryear
{Passy et al.}{2012}]{passy12} 
Passy, J.~C., Mac Low, M.-M., De Marco, O., 2012, \apj, 759, L30




\bibitem[\protect\citeauthoryear
{Qian et al.}{2009}]{qian09} 
Qian, S.-B., Zhu, L.-Y., Zola, S., et al. 2009, \apj, 695, L163

\bibitem[\protect\citeauthoryear
{Qian et al.}{2012}]{qian12} 
Qian, S.-B., Zhu, L.-Y., Dai, Z.-B., et al. 2012, \apj, 745, L23

\bibitem[\protect\citeauthoryear
{Reed et al.}{2010}]{reed10} 
Reed, M.~D., Kawaler S.~D., \O stensen R.~H., et al., 2010, \mnras, 409, 1496

\bibitem[\protect\citeauthoryear
{Reed et al.}{2011}]{reed11} 
Reed, M.~D., Baran, A., Quint, A.~C., et al., 2011, \mnras, 414, 2885

\bibitem[\protect\citeauthoryear
{Reed et al.}{2012}]{reed12} 
Reed, M.~D., Baran, A., \O stensen R.~H., Telting, J., O'Toole, S.~J., 2012,
\mnras, 427, 1245

\bibitem[\protect\citeauthoryear
{Schaffenroth et al.}{2014}]{schaffenroth14} 
Schaffenroth, V., Geier, S., Heber, U., et al., 2014, \aap, 564, A98

\bibitem[\protect\citeauthoryear
{Schuh et al.}{2014}]{schuh14} 
Schuh, S., Silvotti, R., Lutz, R., Kim, S.-L., 2014, ASP Conf. Series, 481, 3

\bibitem[\protect\citeauthoryear
{Shibahashi \& Kurtz}{2012}]{shibahashi12} 
Shibahashi H., Kurtz, D.~W., 2012, \mnras, 422, 738

\bibitem[\protect\citeauthoryear
{Silvotti et al.}{2007}]{silvotti07} 
Silvotti, R., Schuh, S., Janulis, R., et al. 2007, Nature, 449, 189


\bibitem[\protect\citeauthoryear
{Silvotti et al.}{2014}]{silvotti14} 
Silvotti, R., {\O}stensen, R.~H., Telting, J.~H.., Lovis, C., 2014,
ASP Conf. Series, 481, 13

\bibitem[\protect\citeauthoryear
{Soker}{1998}]{soker98} 
Soker, N., 1998, \aj, 116, 1308

\bibitem[\protect\citeauthoryear
{Spiegel}{2012}]{spiegel12} 
Spiegel D.~S., 2012, Proc. of conf. on Planets Around Stellar Remnants
(arXiv:1208.2276)

\bibitem[\protect\citeauthoryear
{Telting et al.}{2012}]{telting12}
Telting J.~H., {\O}stensen R.~H., Baran, A.~S., et al., 2012, \aap, 544, A1

\bibitem[\protect\citeauthoryear
{Valencia et al.}{2010}]{valencial10}
Valencia, D., Ikoma, M., Guillot, T., Nettelmann, N., 2010, \aap, 516, A20

\bibitem[\protect\citeauthoryear
{Van Grootel et al.}{2010}]{vangrootel10}
Van Grootel, V., Charpinet, S., Fontaine, G., et al., 2010, \apj, 718, L97

\bibitem[\protect\citeauthoryear
{Villaver \& Livio}{2007}]{villaver07} 
Villaver, E., Livio, M., 2007, \apj, 661, 1192

\bibitem[\protect\citeauthoryear
{Villaver \& Livio}{2009}]{villaver09} 
Villaver, E., Livio, M., 2009, \apj, 705, L81

\end{thebibliography}
